\documentclass[
superscriptaddress,
 amsmath,amssymb,
 aps,
pra,
twocolumn,
]{revtex4-2}

\usepackage{graphicx}
\usepackage{dcolumn}
\usepackage{multirow}
\usepackage{amsmath}
\usepackage{amssymb}
\usepackage{euscript}
\usepackage{bm}

\usepackage{xcolor}
\usepackage{lipsum}
\usepackage{hyperref}
\usepackage{qcircuit}
\usepackage{graphicx}
\graphicspath{{figures/}}
\usepackage{dcolumn}
\usepackage{bm}
\usepackage{bbold}
\usepackage[utf8]{inputenc}
\usepackage[T1]{fontenc}
\usepackage{mathptmx}
\usepackage{mathtools}
\usepackage{xcolor}
\usepackage{tikz}
\usepackage{braket}
\usepackage{wrapfig}

\definecolor{ao(english)}{rgb}{0.0, 0.5, 0.0}

\usepackage[english]{babel}

\begin{document}

\preprint{AIP/123-QED}


\title[TDHF-YB]{Quantum time dynamics mediated by the Yang-Baxter equation and artificial neural networks}


\author{Sahil Gulania}
\email{sgulania@anl.gov}
\affiliation{Mathematics and Computer Science, Argonne National Laboratory, Lemont, Illinois 60439, USA}

\author{Yuri Alexeev}
\email{yalexeev@nvidia.com}
\affiliation{Computational Science Division, Argonne National Laboratory, Lemont, Illinois 60439, USA}
\affiliation{Present Address: NVIDIA Corporation, Santa Clara, California 95051, USA}

\author{Stephen K. Gray}
\email{gray@anl.gov}
\affiliation{Center for Nanoscale Materials, Argonne National Laboratory, Lemont, Illinois 60439, USA}

\author{Bo Peng}
\email{peng398@pnnl.gov}
\affiliation{Physical and Computational Sciences Directorate, Pacific Northwest National Laboratory, Richland, Washington 99352, USA}

\author{Niranjan Govind}
\email{niri.govind@pnnl.gov}
\affiliation{Physical and Computational Sciences Directorate, Pacific Northwest National Laboratory, Richland, Washington 99352, USA}
\affiliation{Department of Chemistry, University of Washington, Seattle, WA 98195}

\date{\today}
\begin{abstract}
Quantum computing shows great potential, but errors pose a significant challenge. This study explores new strategies for mitigating quantum errors using artificial neural networks (ANN) and the Yang-Baxter equation (YBE). Unlike traditional error mitigation methods, which are computationally intensive, we investigate artificial error mitigation. We developed a novel method that combines ANN for noise mitigation combined with the YBE to generate noisy data. This approach effectively reduces noise in quantum simulations, enhancing the accuracy of the results. The YBE rigorously preserves quantum correlations and symmetries in spin chain simulations in certain classes of integrable lattice models, enabling effective compression of quantum circuits while retaining linear scalability with the number of qubits. This compression facilitates both full and partial implementations, allowing the generation of noisy quantum data on hardware alongside noiseless simulations using classical platforms. By introducing controlled noise through the YBE, we enhance the dataset for error mitigation. We train an ANN model on partial data from quantum simulations, demonstrating its effectiveness in mitigating errors in time-evolving quantum states, providing a scalable framework to enhance quantum computation fidelity, particularly in noisy intermediate-scale quantum (NISQ) systems. We demonstrate the efficacy of this approach by performing quantum time dynamics simulations using the Heisenberg XY Hamiltonian on real quantum devices.

\end{abstract}
\maketitle

\section{Introduction}
In the realm of quantum computing, the tantalizing promise of unprecedented computational power and groundbreaking advancements in various fields has captured the imagination of scientists and innovators worldwide. However, harnessing the full potential of quantum computers is not without its challenges, and one of the most formidable obstacles is the omnipresent issue of quantum errors~\cite{lidar2013quantum}. These errors, arising from the inherently probabilistic and fragile nature of quantum bits or qubits, threaten the accuracy and reliability of quantum computations. To overcome this hurdle, researchers have been fervently exploring strategies for quantum error mitigation (QEM)~\cite{endo2018practical,endo2021hybrid,lowe2021unified,temme2017error,huo2022dual,nation2021scalable}, and among them, a particularly promising approach has emerged: artificial error mitigation (AEM)~\cite{kim2020quantum,kim2022quantum,basu2022qer,kusyk2021survey}. AEM represents a symbiotic marriage between classical and quantum computing, capitalizing on the strengths of each to combat the weaknesses of the other. In this approach, classical algorithms are employed to analyze and model the errors that occur during quantum computations. These error models are then used to guide the application of corrective operations on quantum states, effectively reducing the impact of errors and enhancing the reliability of quantum results. 

At its core, QEM seeks to rectify the deleterious effects of noise and errors in quantum computations. These errors can stem from a multitude of sources, such as imperfect hardware~\cite{kapit2016hardware,battistel2021hardware}, environmental interference~\cite{knill2000theory,sawaya2016error}, or the intrinsic characteristics of quantum bits~\cite{knill1997theory}. Canonical error correction techniques involve error-correcting codes and fault-tolerant quantum circuits~\cite{shor1996fault}, which come at a substantial cost in terms of qubit overhead and computational resources. To alleviate such issues,  artificial error mitigation~\cite{kim2020}, a novel paradigm that leverages the power of classical computation, can be used.
QEM is a vast and rapidly evolving field. Due to its complexity and breadth, this section will focus on highlighting only some of the major contributions, particularly those integrating artificial intelligence (AI) for QEM. Readers are encouraged to consult comprehensive reviews for a more detailed exploration of the QEM subject ~\cite{cai2023quantum,cai2021practical,qin2022overview}.

Using AI for QEM is this field's most recent and exciting development.
Incorporating AI into QEM has shown significant advantages to achieve, or in some cases even exceed, the accuracy of traditional QEM techniques. In a recent  review~\cite{liao2023machine}, the authors performed a comprehensive evaluation that covered a range of AI techniques — such as linear regression~\cite{czarnik2021}, random forests, multi-layer perceptrons~\cite{kim2020}, and graph neural networks~\cite{ying2018,reiser2022} — applied across various types of quantum circuits and various quantum devices.

There are a few other ways in which advanced AI techniques can be applied to QEM. For example, it can be used to adjust probabilities in computational measurements~\cite{smith2020quantum_error_mitigation}. A neural network-based methodology was demonstrated for accurately extracting the noise spectrum from qubits, significantly improving existing techniques~\cite{chen2019neural_network_noise_spectrum}. Machine learning (ML) models can be used to predict near-noise-free expectation values from noisy quantum processing units (QPU)~\cite{wang2021ml_expectation_values}. Another approach is to use a data augmentation-empowered neural model for error mitigation (DAEM)~\cite{liao2023flexible}.

In this article we merge ideas from ML methods such as artificial neural networks (ANN)~\cite{kim2020} and zero-noise extrapolation (ZNE). An issue with the ZNE method is the generation of noisy data using various techniques which are not feasible options for large quantum circuits and large qubit systems. The error accumulated by unitary folding can lead to extra errors, leading to unwanted results. The other issue with quantum time dynamics simulations is to perform the ZNE for each time-step. This incurs significant overhead. Therefore, we use ANNs to learn on few time steps and correct the rest of the dynamics. Specifically, an ANN trained on 30 random samples in time of a 100 time step simulation successfully denoised the remaining steps. Our technique takes advantage of circuit compression using the YBE, which preserves quantum correlations and symmetries in spin chain simulations in certain classes of integrable models, enabling effective compression of quantum circuits while retaining linear scalability with the number of qubits~\cite{bassman2021constantdepth,pengYBE2023,kokcu2022algebraic}. This allows us to control the circuit depth and generate extra noise data without introducing other numerical errors. 

Integrable Hamiltonian problems\cite{sutherland2004beautiful} play a very useful role in quantum computing, such as adiabatic quantum optimization, by providing analytically tractable model Hamiltonians that act as starting points for exploring more complex problems. A notable example is the Ising formulation of various NP-hard problems~\cite{10.3389/fphy.2014.00005}. In this context, our overarching objective is to leverage the algebraic structure of unitary transformations on $2^n$-dimensional Hilbert spaces, along with their approximations, to achieve quantum utility$—$an essential milestone on the path to quantum advantage$—$in solving these problems using both qubit-based and higher-dimensional quantum computing architectures~\cite{pengYBE2023,qutrit_peng}. Additionally, since the exact solutions are known, integrable  models are invaluable for testing quantum simulators and assessing the efficiency of quantum hardware.

The rest of this article is organized as follows. First, we introduce the fundamental concepts of quantum error mitigation, discussing the challenges posed by errors in quantum computations and introducing traditional methods such as zero noise extrapolation (ZNE) and learning-based error mitigation. We then delve into the YBE and its role in compressing time dynamics simulations, providing a constant-depth circuit for certain classes of lattice models. The synergy of YBE with artificial neural networks (ANN) is explored, highlighting the potential for effective error mitigation. We then introduce learning-based error mitigation, showcasing its adaptability and effectiveness, especially in scenarios involving numerous qubits and substantial circuit depths. The learning curve analysis provides insights into the relationship between the amount of training data and the accuracy of the regression model. Finally, we present the outcomes of our study, demonstrating the application of the ANN model in mitigating errors during time dynamics simulations of spin chains. The comparison of raw results, fully compressed circuits, partially compressed circuits using YBE, and the ANN-mitigated values illustrates the effectiveness of our error mitigation approach.

\section{Quantum error mitigation}
In this section we will briefly introduce certain key concepts
and terminology that are common to all the quantum error mitigation (QEM) methods~\cite{endo2018practical,strikis2021learning,takagi2022fundamental}. QEM is an essential facet of quantum computing, addressing the intrinsic vulnerability of qubits to various sources of noise and imperfections. As quantum computers continue to evolve and scale up, the issue of quantum errors become increasingly critical. QEM seeks to understand, quantify, and ultimately mitigate the impact of these errors, enabling more dependable quantum computations. Some approaches include zero noise extrapolation (ZNE)~\cite{he2020zero}, probabilistic error cancellation (PEC)\cite{mari2021extending}, measurement error mitigation (MEC)~\cite{funcke2022measurement}, and approaches like virtual distillation and error suppression by derangement to name a few. We will discuss ZNE in detail as we use this approach in our method. For detailed discussion, we refer the interested reader to more comprehensive reviews on the topic~\cite{cai2023quantum}.

\subsection*{Zero Noise Extrapolation}

Zero noise extrapolation (ZNE) serves as an error mitigation method employed to predict the noiseless expectation value of an observable by extrapolating from a series of expectation values calculated at various noise levels~\cite{he2020zero}. This methodology involves a two-step process. The first step involves deliberately amplifying noise, and various methods can be employed for this purpose. Techniques such as pulse -stretching allow the elevation of the noise level in quantum computations. Similarly, at the gate level, approaches like unitary folding or identity insertion scaling can achieve comparable outcomes.
The second step  entails extrapolating to the noiseless limit. This is achieved by fitting a curve, commonly referred to as an extrapolation model, to the expectation values recorded at different noise levels. The goal is to extrapolate and obtain the expectation value in the absence of noise.

\subsubsection*{Scaling Noise}
\begin{figure}[!ht]
    \centering
    \includegraphics[width=\linewidth]{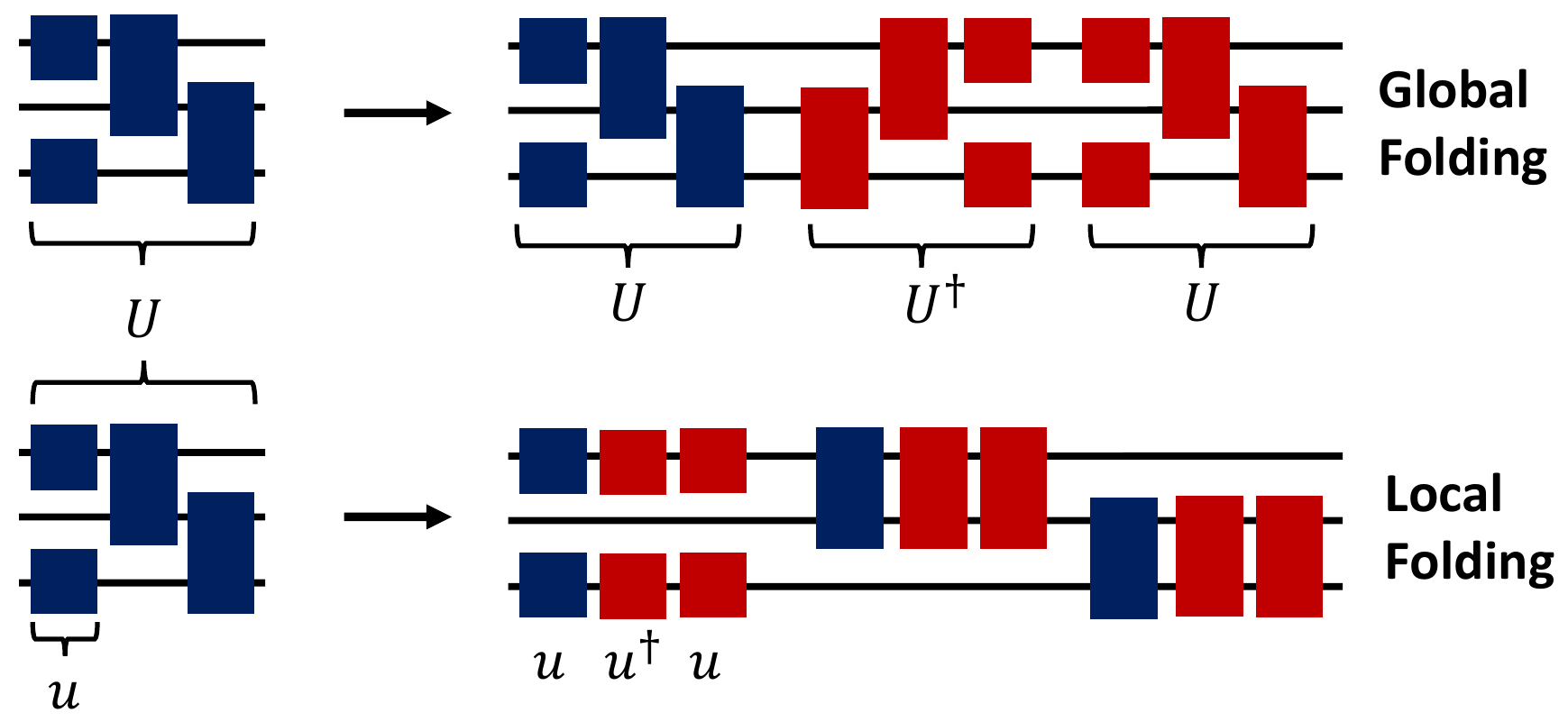}
    \caption{Different variation of performing folding of gates. Top diagram shows the global folding where the whole unitary, $U$ is folded. The bottom diagram shows local folding where each part of whole unitary is folded individually.}
    \label{fig:folding}
\end{figure}

\begin{figure}[!ht]
    \centering
    \includegraphics[width=\linewidth]{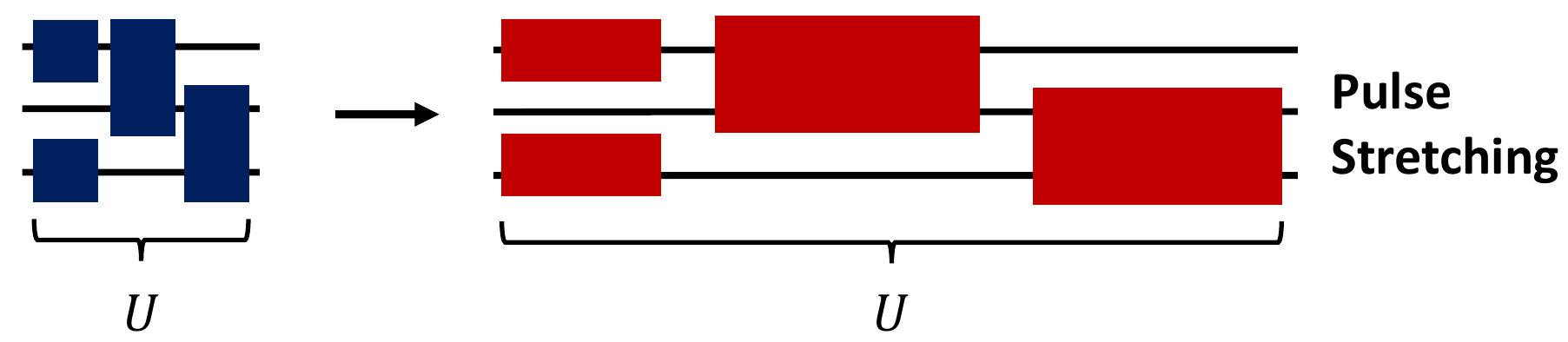}
    \caption{Pulse stretching to increase noise in a physical device}
    \label{fig:pulse}
\end{figure}

An approach to heighten the noise level within a circuit at the gate level involves deliberately enhancing its depth. This can be achieved through either unitary folding or identity scaling. During unitary folding, a mapping process is executed as
\begin{equation}
U \rightarrow UU^{\dagger}U    
\end{equation}
This mapping can be implemented on a global scale or applied locally, as illustrated in the Figure \ref{fig:folding}.
One can also introduce the noise by adding identity operators as
\begin{equation}
    U \rightarrow I. U
\end{equation}
The sole distinction between folding and identity insertion lies in the fact that, rather than scaling gate noise, the introduction of an identity gate serves to extend the waiting period subsequent to the execution of each circuit layer. This extension enables qubits to engage with the environment through a noisy process, potentially undergoing decoherence if the interaction between the system and its environment is substantial. 
Both the techniques are gate level method to introduce noise. It can also be achieved by pulse-stretching method. The noise of the device can be modified by increasing the time over which pulses are implemented, which is shown in Figure \ref{fig:pulse}.

\subsubsection*{Extrapolation}
The fundamental idea behind the ZNE technique is to remove the noise very specific to a given circuit. Let, $\gamma$ be a parameter which quantifies the noise for a quantum circuit and let $\gamma^{'}$ be the parameter for noise in the scaled quantum circuit. ZNE assumes that $\gamma^{'}$ can be equated to $\gamma$ linealry as
\begin{equation}
    \gamma^{'} = \lambda \gamma
\end{equation}
For $\lambda = 1$, the input quantum circuit remains unchanged and the noise level $\gamma = \gamma^{'}$, which is same as the noise of the device without any scaling.
In terms of quantum state $\rho$ and expectation value $E$, one can formulate the same relation with noise level. Let $\rho(\gamma^{'})$ be the quantum state prepared by scaled quantum circuit. One can compute the expectation value of an observable $M$ as
\begin{equation}
    \braket{E(\lambda)} = Tr[\rho(\gamma^{'})M] = Tr[\rho(\lambda \gamma)M]
\end{equation}
The scaled quantum circuit allows to compute different expectation value dependent on $\lambda$. The final aim to compute $E(\lambda=0)$ which corresponds to noiseless expectation value. In practice the $E(\lambda)$ is treated as a function and act as an input to extrapolation model to predict the zero-noise limit ($\lambda = 0$). Various selections for the extrapolation model result in different extrapolations. Common options for the extrapolation model include a linear function, a polynomial, and an exponential function.

\section{Yang-Baxter equation}

The Yang--Baxter equation~\cite{jimbo1989introduction,Yang-Baxter-BBC} was introduced independently in theoretical physics by Yang~\cite{yang1967some} in the late 1960s and by Baxter \cite{baxter1972partition} in statistical mechanics in the early 1970s. This relation has also received much attention in many areas of theoretical physics, classification of knots, scattering of subatomic particles, nuclear magnetic resonance, and ultracold atoms and, more recently, in quantum information science \cite{ge2016YBE,nayak2008non,kauffman2010topological,zhang2013integrable,vind2016experimental,batchelor2016yang}. 

The YBE connection to quantum computing originates from investigating  the relationship between topological entanglement, quantum entanglement, and quantum computational universality. Of particular interest is  how the global topological relationship in spaces (e.g., knotting and linking) corresponds to the entangled quantum states and how the CNOT gate, for instance, can in turn be replaced by another unitary gate $R$ to maintain universality. It turns out these unitary $R$ gates, which serve to maintain the universality of quantum computation and also serve as solutions for the condition of topological braiding, are unitary solutions to the YBEs~\cite{baxter2016exactly}. Put another way, the YBE relation is a consistency or exchange condition that allows one to factorize the interactions of three bodies into a sequence of pairwise interactions under certain conditions.

\begin{figure}[!ht]
    \centering
    \includegraphics[width=\linewidth]{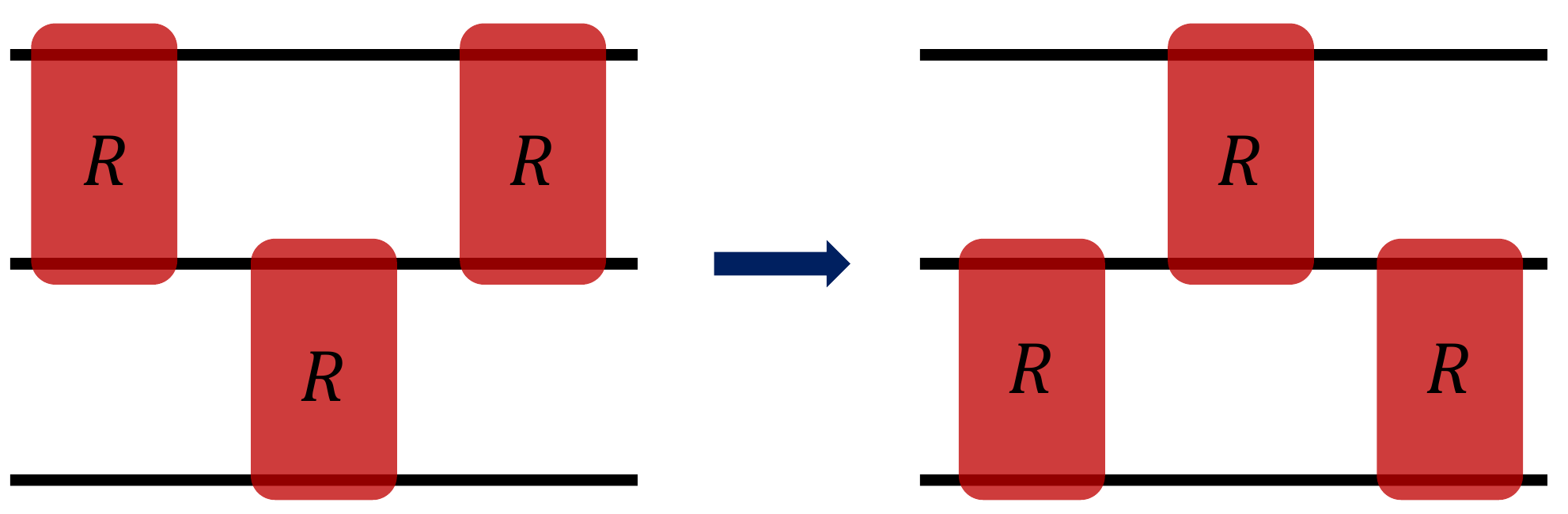}
    \caption{Quantum circuit representation of the YBE for three qubits.}
    \label{fig:ybe}
\end{figure}

This can be formally written as 
\begin{equation}
    (R\otimes I)(I\otimes R)(R\otimes I) = (I\otimes R)(R\otimes I)(I\otimes R) 
    \label{eq:YBE}
\end{equation}
where $R$ is an operator acting on two qubits and $I$ is the identity operator acting on single qubit. The 
$R$ operator is also a linear mapping $R: V\otimes V \rightarrow V\otimes V$ defined as a twofold tensor product generalizing the permutation of vector space $V$. This relation also yields a sufficiency condition for quantum integrability in one-dimensional quantum systems and provides a systematic approach to construct integrable models. A diagrammatic representation of YBE in the gate set and quantum circuit form is shown in Figure \ref{fig:ybe}.

Recently, a novel perspective on this equation has emerged in the context of the quantum time dynamics of lattice models~\cite{pengYBE2023,qutrit_peng}. Specifically, under the Trotter approximation~\cite{nielsen2010quantum}, the YBE enables the compression of any arbitrary time step in certain lattice models into a circuit of finite depth without compromising accuracy. 
As a result of this transformation, an additional symmetry, referred to as mirror symmetry, can be deduced in the circuit~\cite{pengYBE2023,bassman2021constantdepth,kokcu2022algebraic,gulaniareflection}. Furthermore, the two-qubit operations adhere to the merge identity. The combination of new relations allow one to obtain a constant depth circuit for certain classes of quantum operations~\cite{pengYBE2023}.

\subsection*{Compression Algorithm}
The YBE compression algorithm is composed of two parts. The first part is the merge identity, where the two different two qubit operators ($R_{1},R_{2}$) acting on a pair can be compressed as a single two qubit operators ($R_{3}$). 
\begin{equation}
R_{1} \times R_{2} = R_{3} \hspace{2cm} \forall R_{i} \in \mathbb{YBE}
\end{equation}
The second part is reflection symmetry. One can use the YBE relation recursively to show the existence of reflection symmetry for N-qubits operations\cite{pengYBE2023}. Reflection symmetry, in conjunction with the merge identity, enables the compression of N alternate layers of gates into N/2 alternate layers for N qubits. Fig.\ref{fig:compression_algo_2} illustrates this concept for four qubits, where a third alternate layer is merged with the preceding two layers. Consequently, any number of alternate layers can be condensed into two alternate layers~\cite{pengYBE2023,gulaniareflection}.
\begin{figure}[!h]
    \includegraphics[width=\linewidth]{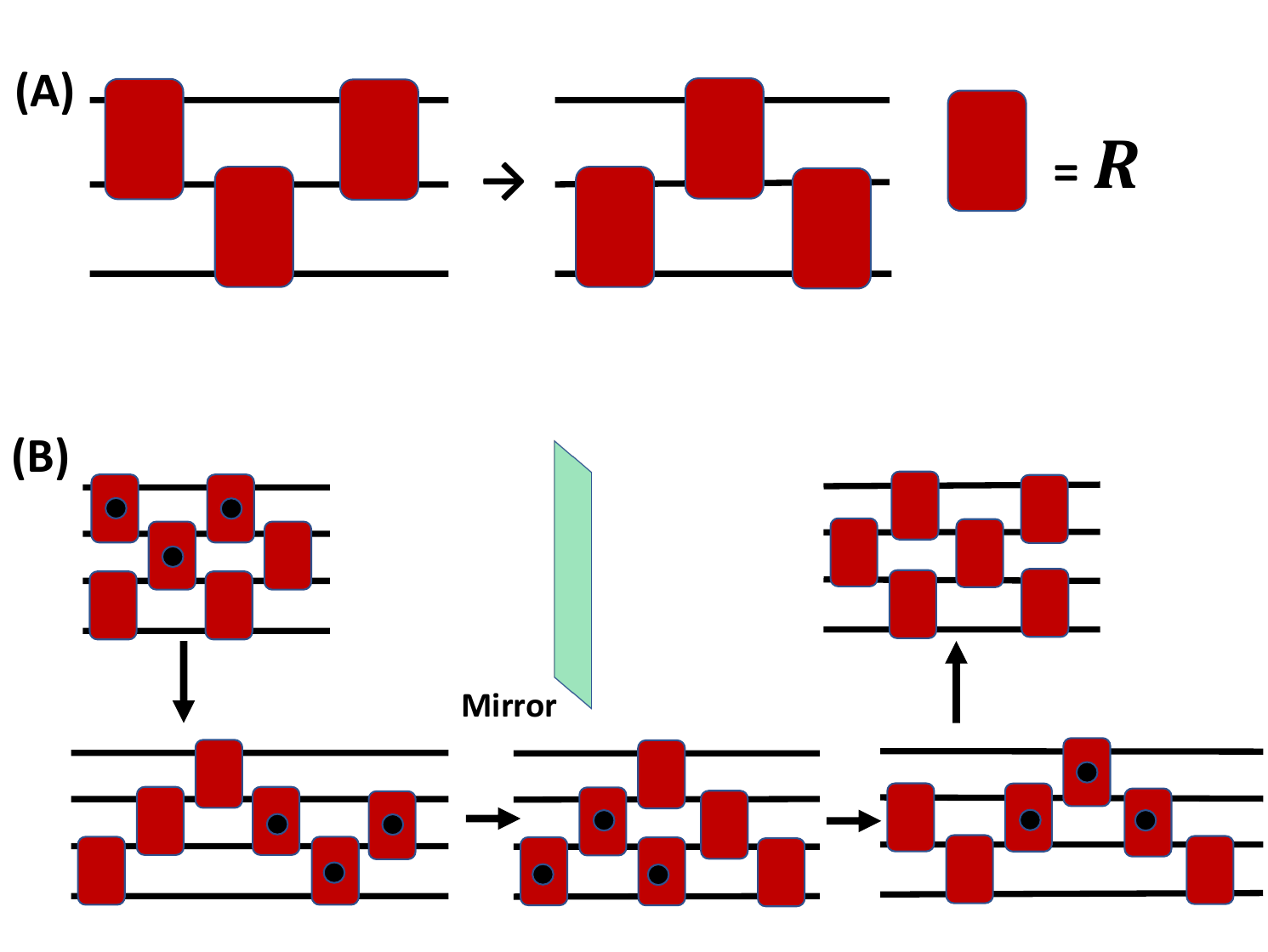}
    \caption{(A) Quantum circuit representation of the YBE for three qubits. (B) Reflection symmetry is achieved by using the YBE four times on four qubits (action of YBE on which triplets is shown by black dots)}
    \label{fig:compression_algo_1}
\end{figure}

\begin{figure}
    \includegraphics[width=\linewidth]{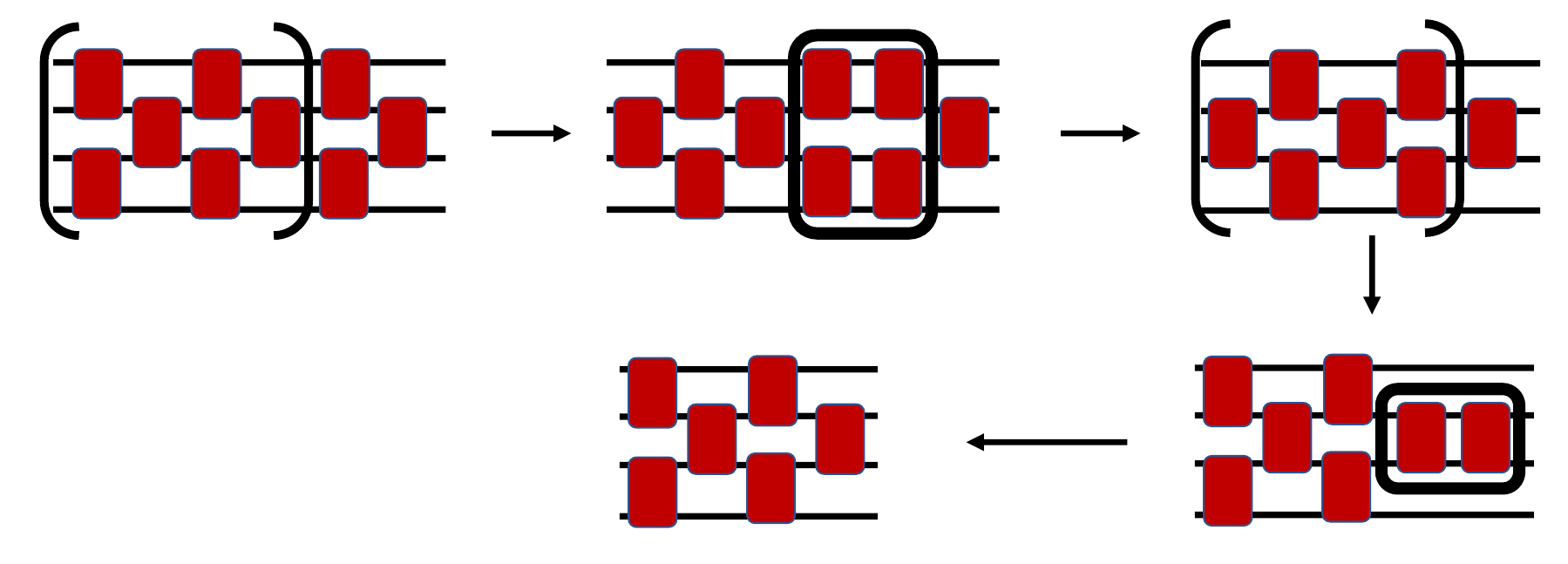}
    \caption{Compression scheme for 4 qubits. Reflection symmetry exists with two layers of alternative gates. Addition of a third layer can be absorbed into the two layers by recursive usage of reflection symmetry (red bracket) via the YBE and merge identity.}
    \label{fig:compression_algo_2}
\end{figure}

\section{Learning Based Error Mitigation}
\begin{figure}[!ht]
    \centering
    \includegraphics[width=\linewidth]{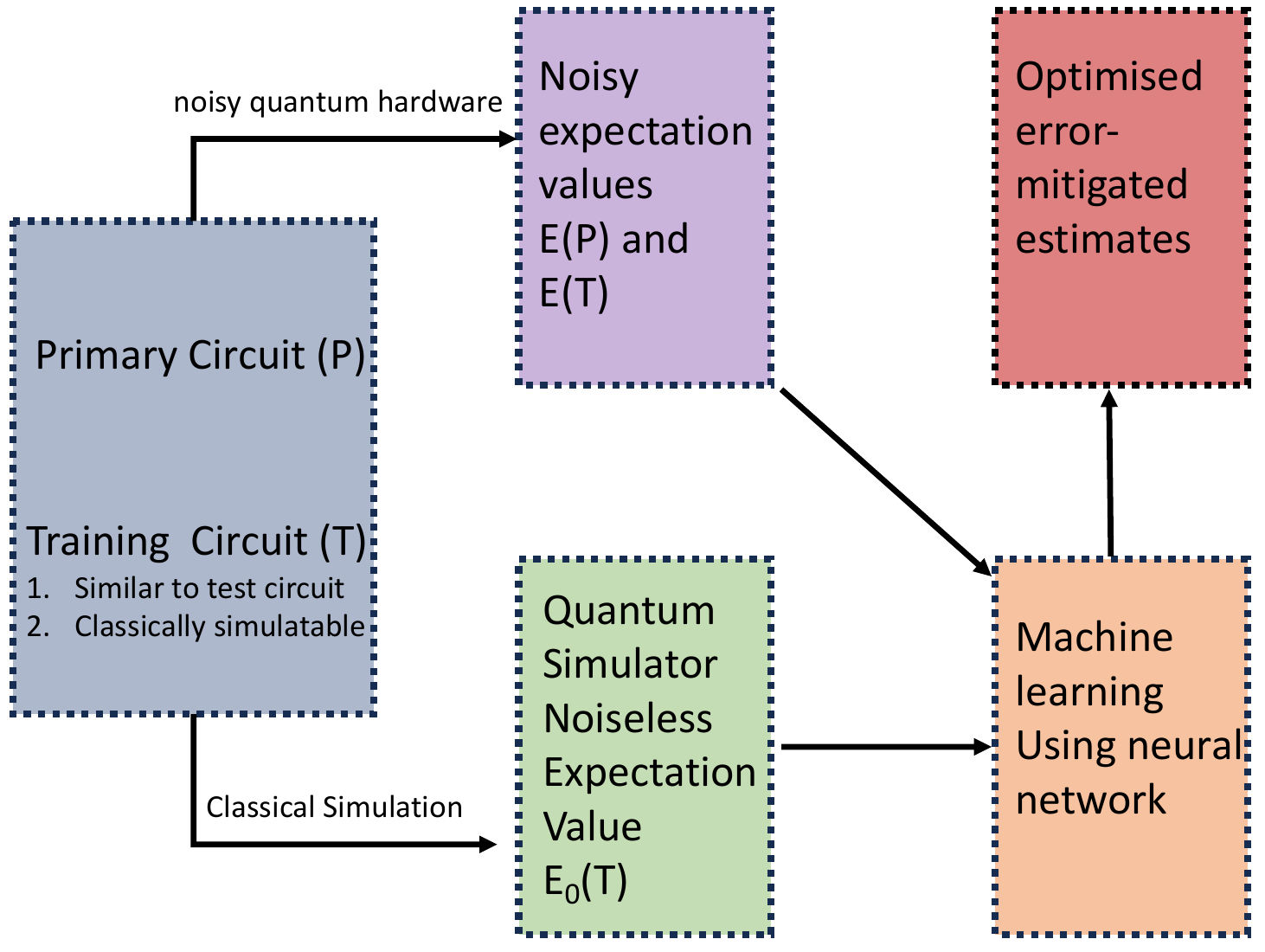}
    \caption{Diagram showing the process of learning-based quantum error mitigation.}
    \label{fig:learning}
\end{figure}

A recently proposed method, known as learning-based error mitigation is discussed in this section. This approach involves understanding the impact of noise by examining classically simulatable quantum circuits closely resembling the non-simulatable circuits of interest. In many instances, this approach evaluates the expectation value of an observable for specific training circuits on a quantum device, utilizing training data obtained through classical means. The training data is then analyzed using an ansatz, which captures the relationship between noisy and noiseless expectation values. The resulting ansatz is applied to correct the noisy expectation value for the target circuit.  
A diagram showing the process of learning based quantum error mitigation is shown in Figure \ref{fig:learning}. 

Learning-based error mitigation has proven effective, particularly in applications involving quantum circuits with numerous qubits and substantial depths~\cite{bultrini2023unifying}. This approach is highly adaptable, allowing for the improvement of mitigation quality by expanding training data to encompass the effects of varying noise strengths on observable expectation values. These characteristics position learning-based error mitigation as a promising solution for addressing errors in near-term quantum advantage applications.

In the proposed methodology, the neural network is provided with multiple inputs and learns to remove the quantum noise. The inputs are as follows: 
\begin{enumerate}
\item Staggered magnetization from a partially compressed quantum circuit using a real quantum device.
\item Staggered magnetization from a fully compressed quantum circuit using a real quantum device.
\item Zero noise - staggered magnetization from a classical device.
\item Time step parameter.
\item Spin system – number of spins.
\end{enumerate}
The main ingredient in this learning process is the partial noise addition from a partial compressed circuit. As YBE only scales noise while maintaining its form. The intrinsic idea behind using partial compressions from YBE lies in gates used in quantum time dynamics of the relevant spin chain model. Partial compression allows one to reach the same point in the Hilbert space with some delay compared with full compression.

\section{Experiment}
Our experiment involves a multifaceted approach encompassing quantum time dynamics simulations, error mitigation techniques, and machine learning approaches introduced in the previous sections. We have investigated the time dynamics of the XY Hamiltonian, a subclass of the general Heisenberg Hamiltonian. The Heisenberg Hamiltonian \cite{fazekas1999lecture, skomski2008simple, pires2021theoretical} has been widely used to study magnetic systems, where the magnetic spins are treated quantum mechanically where the Hamiltonian with nearest neighbor interactions can be written as
\begin{equation}
    \hat{H} = 
    -\sum_{\alpha}\{J_{\alpha}\sum_{i=1}^{N-1} \sigma_{i}^{\alpha}\otimes \sigma_{i+1}^{\alpha}\}, 
\label{eq:Heisenberg}
\end{equation}

\begin{figure}[!h]
    \includegraphics[width=\columnwidth]{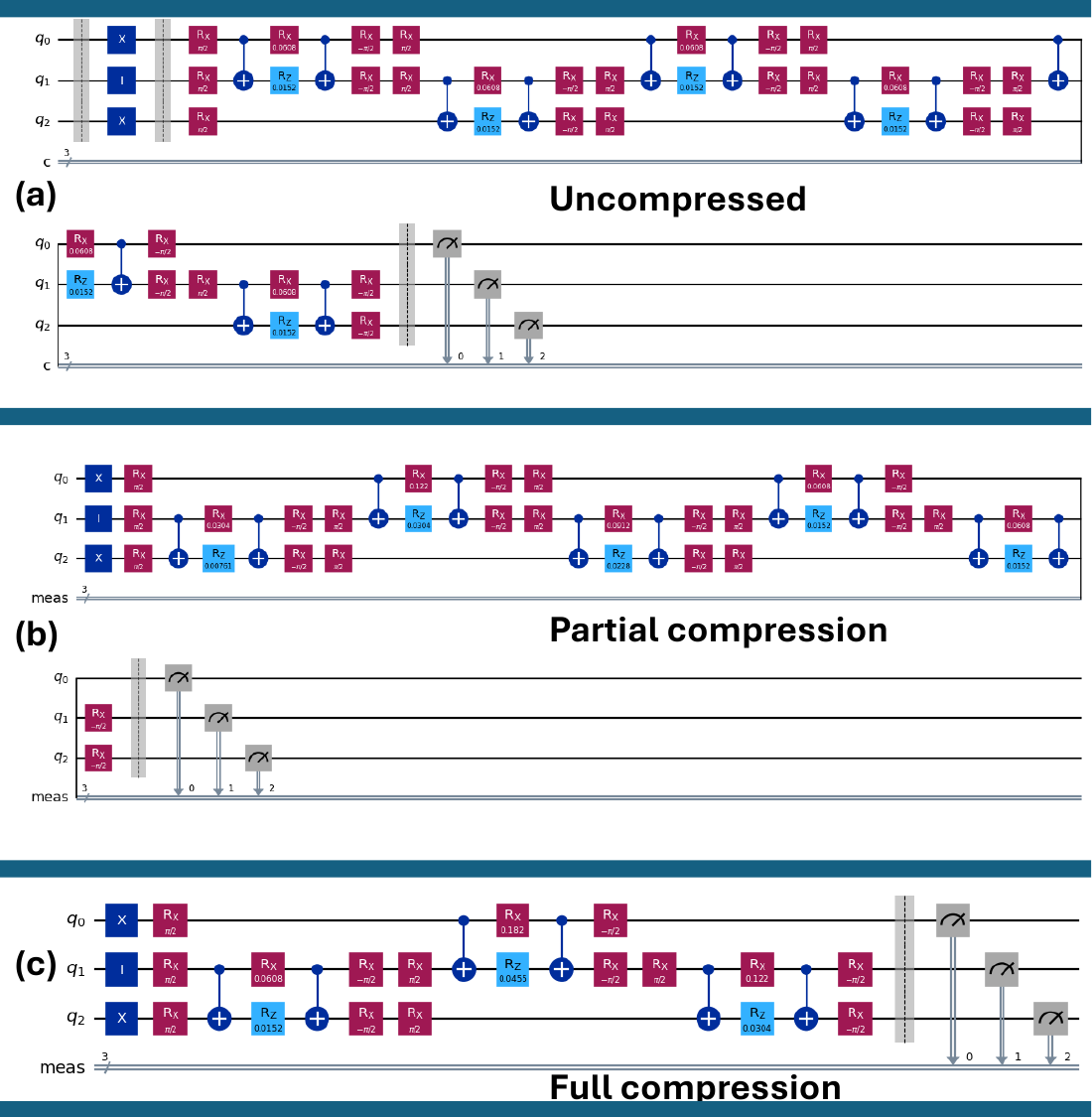}
    \caption{Full quantum circuit for a 3-spin XY model ($J_x = -0.8$
    \& $J_y = 0.2$) with a time step of 0.025 units. (a) Quantum circuit without any compression for three Trotter steps. It has a total of 12 CNOTs. (b) Quantum circuit with partial compression using YBE compression scheme for three Trotter steps. It has a total of 8 CNOTs. (c) Quantum circuit with full compression using YBE compression scheme for three trotter steps. It has a total of 6 CNOTs.}
    \label{fig:compression comparison}
\end{figure}

\noindent{where} $\alpha$ sums over $\{x,y,z\}$, the coupling parameter $J_{\alpha}$ denotes the exchange interaction between nearest-neighbour spins along the $\alpha-$direction, and $\sigma^{\alpha}
_{i}$ is the $\alpha$-Pauli operator on the $i$th
spin. For the XY Hamiltonian\cite{lieb1961two,katsura1962statistical,katsura1963statistical}, $J_z = 0$ in Eq. (\ref{eq:Heisenberg}). For the isotropic XY Hamiltonian, $J_x = J_y = J, J_z =0$, while for the anisotropic XY Hamiltonian, $J_x \neq J_y, J_z =0$. Our experiment is as follows: 

\begin{enumerate}
    \item The anisotropic XY Hamiltonian is time evolved for spin chains of different lengths. The staggered magnetization, denoted as $m_s(t)$, is calculated over a time period of 2.5 units, using a Trotter step size of 0.025 units and $m_s(t)$ is given by
    \begin{equation}
         m_s(t) = \frac{1}{N} \sum_{i} (-1)^{i}\braket{\sigma_{z}(t)}. 
    \end{equation}
    \item YBE compression is utilized to construct a constant-depth circuit (Eq. \ref{eq:YBE}) that scales linearly with the number of qubits. 
    \item Partial compression is also used to introduce additional noise into the system, resulting in larger circuits compared to full compression. Fig.(\ref{fig:compression comparison}) shows an example for a 3-spin system. 
    \item Staggered magnetization data obtained from the fully compressed circuit, partially compressed circuit, and noiseless simulator are used to train an ANN model. The ANN is trained on 30 random time steps, and the remaining is used for prediction. Each ANN is trained for different spin-system individually.
     
    \item The results are compared for different spin chains (3-5 spins and 6-10 spins). Raw results from IBM quantum devices with full and partial compression are compared with the ANN values.
    \item The learning curve for the ANN is analyzed to understand the relationship between the number of data points and the accuracy of the model. This provides insights into the performance of the model as more data is used for training.
\end{enumerate}

\section{Results}
Applying Zero-Noise Extrapolation (ZNE) to every time step in a quantum dynamics simulation is highly resource-intensive, as the computational cost grows exponentially with the number of time steps. The evolving quantum state and its interaction with the environment result in a unique noise profile for each step, requiring separate noise extrapolation. This repetitive process places a significant burden on computational resources, making it impractical for large-scale or long-duration simulations.

\begin{figure}[h!]
    \centering   \includegraphics[width=1.0\linewidth]{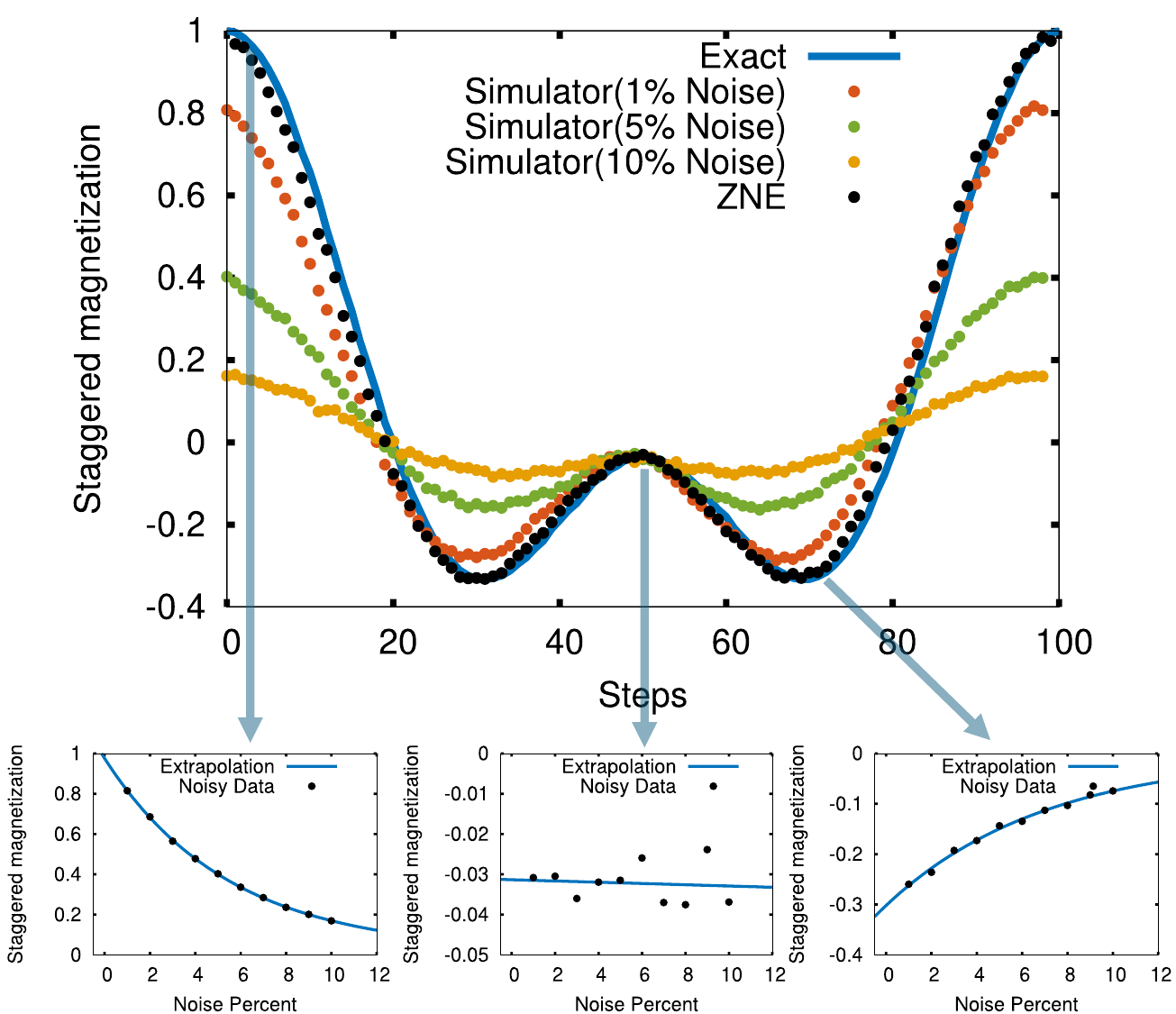}
    \caption{Zero-Noise Extrapolation (ZNE) applied to quantum time dynamics for the XY model over 100 time steps. The main plot shows the exact simulation, noisy simulations with 
    1$\%$, 5$\%$, and 10$\%$ noise, and the ZNE-corrected data. All simulations were performed on a Qiskit simulator. The bottom three graphs display the different functional forms required for fitting the ZNE data at different time-step, highlighting the varying extrapolation behavior across different noise profiles. These results demonstrate the challenges of applying ZNE to time dynamics simulations with varying noise levels at each time step.}
    \label{fig:ZNE}
\end{figure}

To demonstrate these challenges, we conducted a 3-qubit simulation on a noisy simulator, generating noisy data for each time step. ZNE was applied at every step, as shown in Fig. \ref{fig:ZNE}, exposing the limitations of this approach. The need to handle distinct noise profiles for each step leads to substantial computational overhead, restricting the scalability of ZNE for quantum dynamics simulations.

In order to show the novelty of YBE-mediated ANN error mitigation we applied YBE compression to each spin system, enabling the construction of a constant-depth circuit that scales linearly with the number of qubits. Additionally, to introduce extra noise into the system, we utilized partial compression, resulting in larger circuits compared to full compression. All the simulations were run on 
{\textit {ibm-quito}}(3 qubits), {\textit {ibm-manilla}}(4-5 qubits) and
{\textit {ibm-kolkata}}(6-10 qubits) ,  with 100000 shots for noisy quantum data. 

We used the staggered magnetization data obtained from fully compressed circuit, partially compressed circuit and noiseless simulator to train ANN. From the full data point of time-dynamics (100 steps for 3-5 qubits and 50 steps for 6-10 qubits) we used partial data (30 \%) for the training and the rest was used for prediction.
The ANN consists of an input layer with 5 neurons corresponding to the 5 input parameters, followed by two dense (fully connected) layers. The first dense layer processes the input using weighted connections, biases, and an sigmoid activation function, producing an intermediate representation. The second dense layer further processes this representation and outputs a single value, corresponding to your target parameter. During training, the network learns by iteratively minimizing the error between noiseless output from simulator and noisy output from real quantum device.  The process is repeated over multiple epochs to optimize the network's performance.
As shown in the Fig 1 and 2, the ANN is able to perform the mitigation and for every spin chain it is able to push the noisy data closer to the noiseless results. The ANN was implemented as a sequential model using TensorFlow and Keras. The network was trained on 30\% of the available data, with mean squared error (MSE) as the loss function. Optimization was performed using the Adam optimizer. The training process was carried out over multiple epochs, with early stopping criteria to prevent overfitting. The trained model was then applied to predict results for the remaining 70\% of the data.
    \begin{figure}[h!]
        \centering\includegraphics[width=\columnwidth]{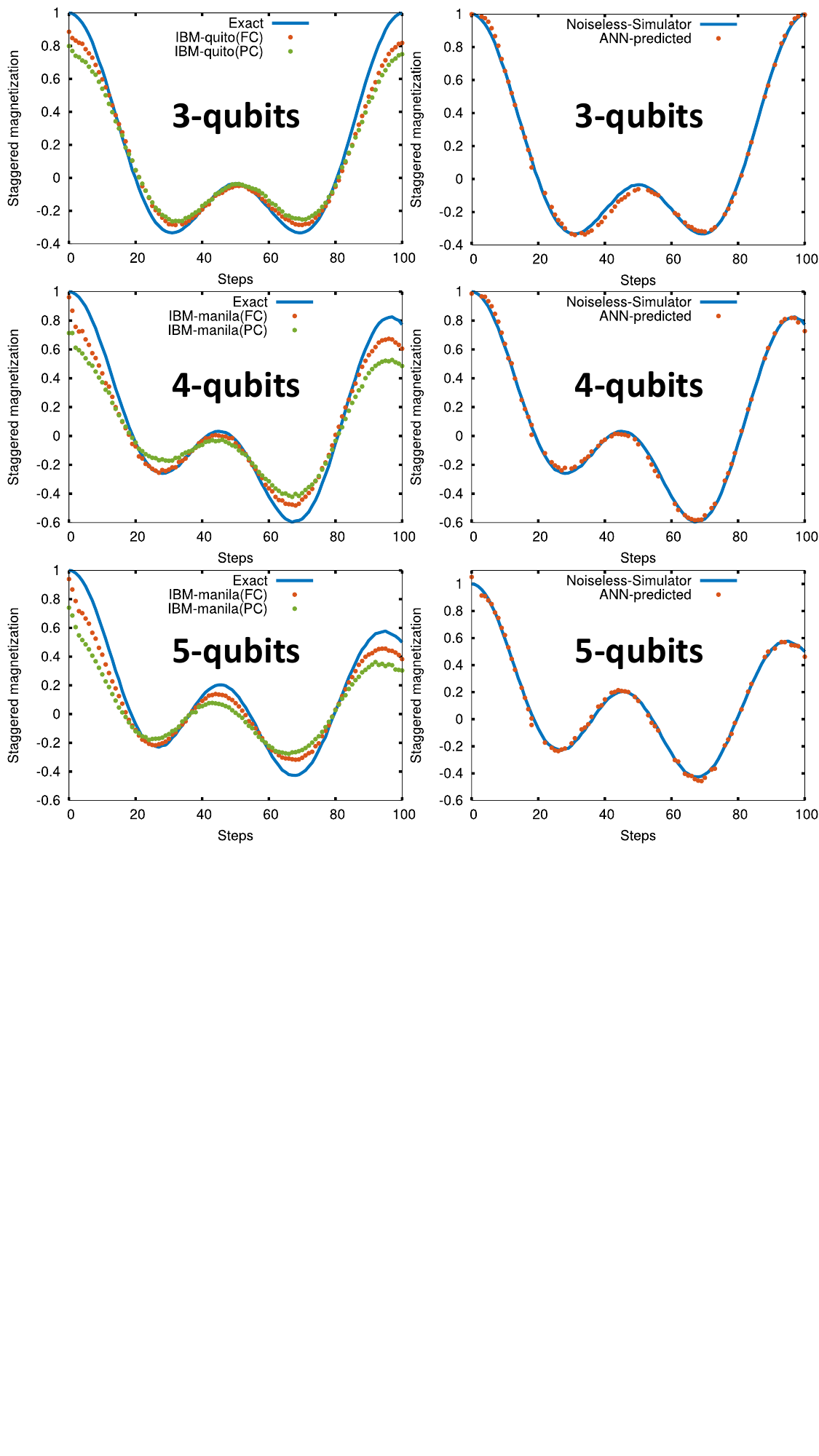}
        \caption{Comparison of results for different spin chains of length 3, 4, and 5. Left columns shows the raw results obtained from the IBM quantum devices with full compression (FC) and partial compression (PC) using YBE. 
        FC corresponds to maximum noise reduced output which YBE can produce. PC corresponds to extra noise output using YBE partially.        
        The right columns shows the ANN mitigated values. ANN was trained on 30 percent of data and was used to predict values for the rest 70 percent}
        \label{fig:ANN_3_5}
    \end{figure}
    
    \begin{figure}[h!]
        \centering\includegraphics[width=\columnwidth]{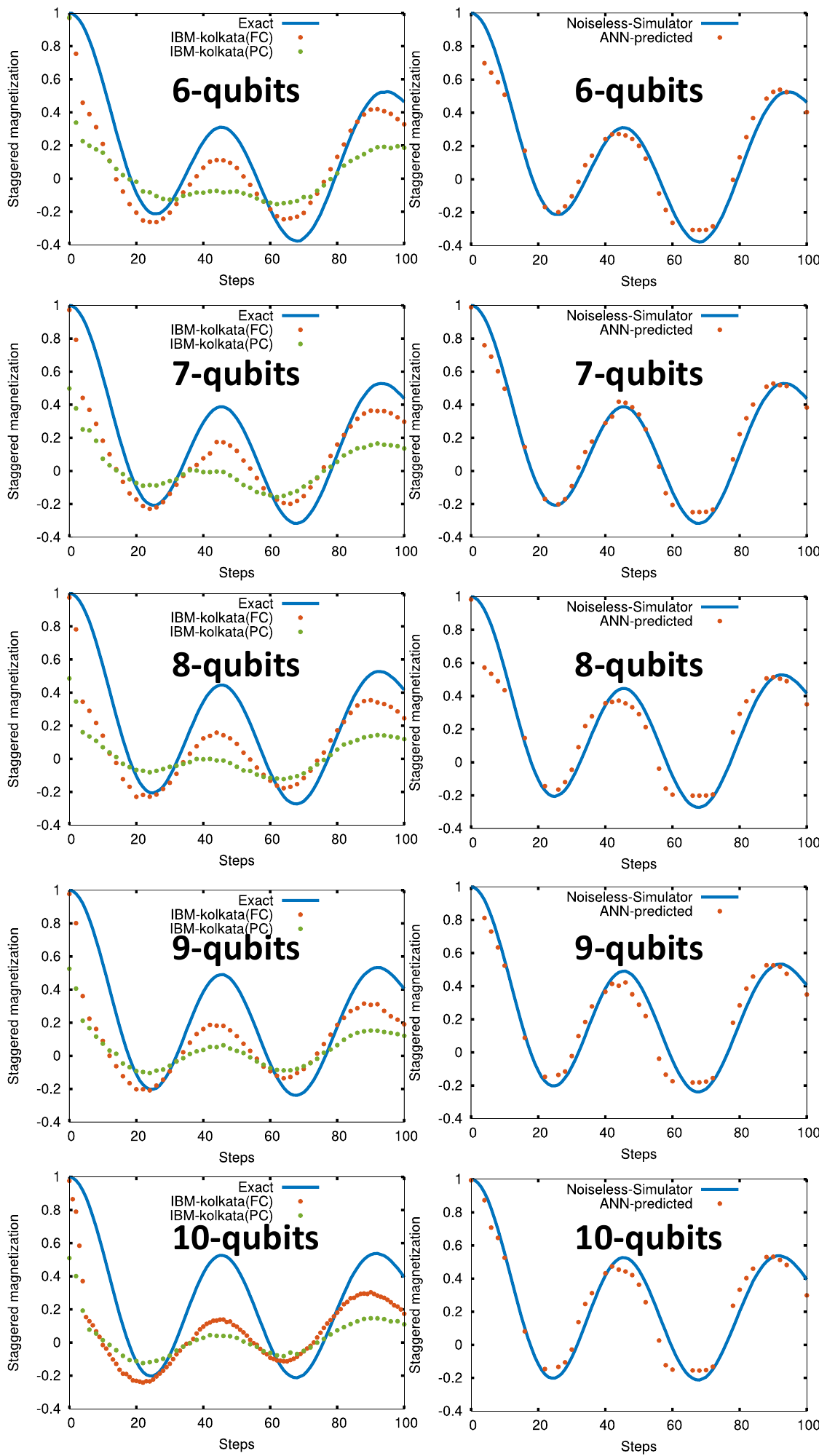}
        \caption{Comparison of results for different spin chains of length 6 to 10 spins. Left columns shows the raw results obtained from the IBM quantum devices with full compression (FC) and partial compression (PC) using YBE. 
        FC corresponds to maximum noise reduced output which YBE can produce. PC corresponds to extra noise output using YBE partially.  The right columns shows the ANN mitigated values. ANN was trained on 30 percent of data and was used to predict values for the rest 70 percent}
        \label{fig:ANN_6_10}
    \end{figure}

\begin{figure}[!ht]
    \centering
    \includegraphics[width=\columnwidth]{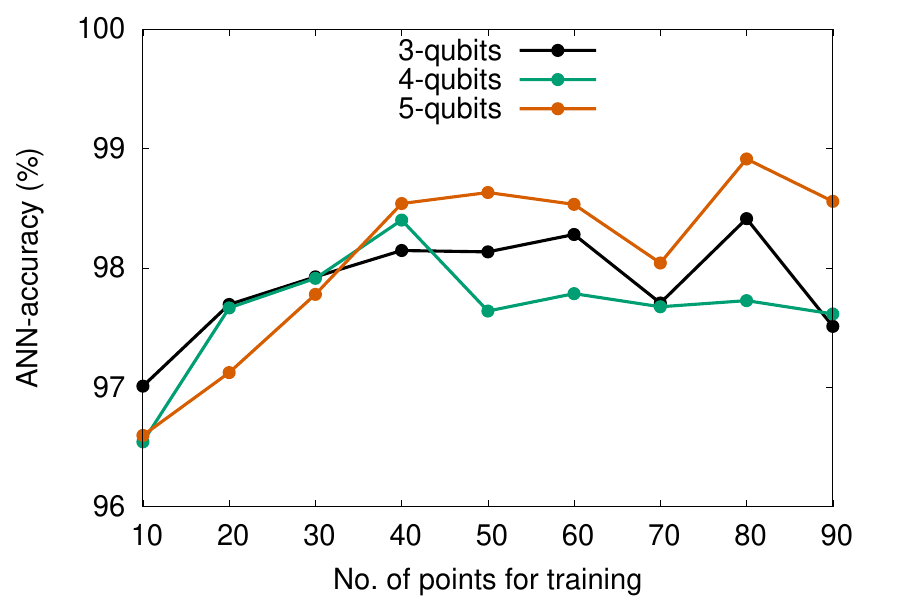}
    \caption{Learning Curve: Impact of varying data points (10 to 90) on Model Accuracy for 3-5 spin system. The total dataset comprises 100 points.}
    \label{fig:3-5_learning}
\end{figure}

The observed learning curve as shown in Fig ~\ref{fig:3-5_learning} suggests insights into the relationship between the number of data points used and the learning performance of regression model as discussed below
\begin{itemize}
    \item The initial steep increase in learning accuracy from 10 to 40 data points indicates that with a small amount of data, the model rapidly improves its understanding and predictive capability. This is typical when a model encounters new information and adjusts its parameters to fit the available data better.
    \item The subsequent flattening of the curve between 40 and 70 data points suggests that adding more data points during this range has diminishing returns in terms of improving the model's accuracy. The model has likely captured the underlying patterns in the data, and additional information doesn't significantly contribute to enhancing its performance.
    \item The oscillations observed from 70 to 90 data points indicate a more complex relationship between the model and the data. It's possible that the model is becoming sensitive to specific data points or noise, leading to fluctuations in accuracy. This behavior may suggest that the model is starting to overfit the training data.
    \item The overall pattern highlights the importance of finding the right balance in the amount of training data. Too few data points may result in underfitting, where the model fails to capture the underlying patterns. However, beyond a certain point, additional data may not significantly improve the model and could even lead to overfitting.
\end{itemize}
\begin{figure}[!ht]
    \centering
    \includegraphics[width=\columnwidth]{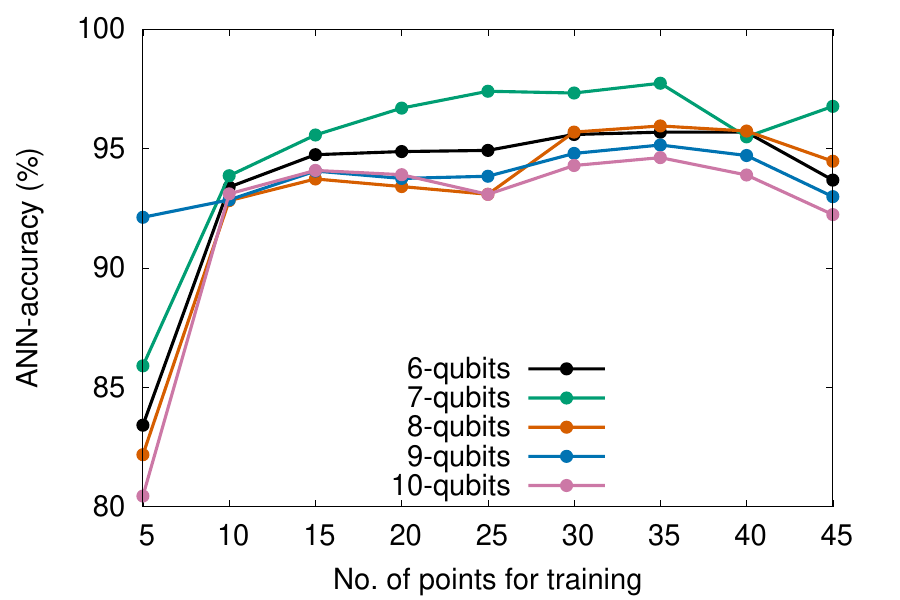}
    \caption{Learning Curve: Impact of varying data points (5 to 45) on Model Accuracy for 6-10 spin system. The total dataset comprises 50 points.}
    \label{fig:6-10_learning}
\end{figure}
The same behavior is observed in the learning curve for 6 to 10 spin chains (as shown in Fig ~\ref{fig:6-10_learning}), which 
mirrors the pattern described earlier. Initially, there's a steep increase in learning accuracy from 5 to 15 points, followed by a plateau from 20 to 35 points, and finally, oscillations in accuracy from 35 to 45 points. This pattern aligns with the principles of model learning and the impact of varying data points~\cite{shalev2014understanding,domingos2012few}. In other words, this is a consequence of overfitting.  

\section{Conclusion and Outlook}
In this study, we demonstrated the effectiveness of an Artificial Neural Network (ANN) in reducing errors during quantum time dynamics simulations. The model learns from partial data, making it well-suited for tracking the evolution of quantum states. Our analysis shows a characteristic learning pattern: accuracy improves rapidly with a small dataset, levels off at intermediate sizes, and fluctuates for larger datasets.

To introduce additional noise, we applied the Yang-Baxter Equation (YBE) compression technique.\cite{pengYBE2023,gulaniareflection} Even when trained on just two noisy data points per time step, the ANN remained effective. By treating each spin’s evolution independently, the model successfully mitigates errors. We are also exploring training the ANN on all spins simultaneously, which could enhance its performance for larger spin chains. This approach is inspired by Gaussian process methods \cite{miller2012bi}. In large-scale simulations, high-performance computing can be used for a limited number of steps to calibrate noise. For instance, out of 100 quantum experiments, only 10 need to be run using resource-intensive quantum computing to learn the noise. Once trained, the neural network can correct noise in the remaining 90 experiments. Our results also show that a simple neural network with just two hidden layers requires only 30

Beyond this, our technique applies to Hamiltonians following Yang-Baxter symmetry, such as mean-field Hamiltonians, expanding its potential for quantum error mitigation and extracting insights from complex quantum systems. Future work will focus on a detailed analysis of how different noise types and levels impact the model's performance using a quantum simulator. Understanding these effects will help refine the model for practical quantum applications.

Second would be extension of the error mitigation technique to perform large-scale spin simulations on actual quantum devices is a crucial step. This involves implementing the model on quantum hardware, where error mitigation becomes essential for accurate results. Studying the effectiveness of the technique in a real-world, noisy quantum environment will be instrumental in advancing quantum computing applications. Our future research will explore alternative learning methods to enhance the accuracy of error mitigation.

Comparing and contrasting the performance of various machine learning algorithms or incorporating hybrid approaches may uncover more effective strategies. Understanding how different learning methods respond to varying numbers of data points can provide guidance for optimizing the error mitigation process. Also, to facilitate broader adoption and application of the error mitigation technique, there is a need to develop a user-friendly software package. This package could encapsulate the methodology, algorithms, and best practices for implementing error mitigation using learning methods. Open-sourcing such a package would contribute to the collaborative advancement of quantum computing research and its practical applications.

Training artificial neural networks (ANNs) for quantum error mitigation (QEM) holds promise, but their applicability to a wide variety of quantum models, particularly many-body systems, remains a challenging task. While ANNs can be trained effectively on random circuits or Clifford gates, scaling this approach to arbitrary quantum circuits is more complex. Specifically, an ANN trained on one type of quantum circuit may not generalize well to another, especially in the case of models with more intricate dynamics, such as those found in many-body quantum systems. Future research is focused on overcoming these limitations, with efforts directed toward developing techniques for training ANNs that can handle a broader range of quantum circuits and effectively mitigate noise across diverse quantum models.

These future studies collectively aim to deepen our understanding of error mitigation in quantum simulations, extend the technique to real-world quantum devices, explore diverse learning methodologies, and provide accessible tools for researchers and practitioners in the field. The outcomes of these endeavors will play a pivotal role in advancing the capabilities and reliability of quantum computing technologies.

\section*{Acknowledgment}
This material is based upon work supported by the U.S. Department of Energy, Office of Science, National Quantum Information Science Research Centers at Argonne National Laboratory and Pacific Northwest National Laboratory (B.~P., N.~G. under FWP 76155). Y.~A. and S.~G also acknowledge support from the U.S. Department of Energy, Ofﬁce of Science, under contract DE-AC02-06CH11357 at Argonne National Laboratory. Work performed at the Center for Nanoscale Materials, a U.S. Department of Energy Office of Science User Facility, was supported by the U.S. DOE, Office of Basic Energy Sciences, under Contract No. DE-AC02-06CH11357. N.~G. and B.~P, also acknowledge the Quantum Algorithms and Architecture for Domain Science Initiative (QuAADS), a Laboratory Directed Research and Development (LDRD) Program at PNNL for work on extensions to time domain simulations of many-body quantum systems. This research used resources of the Oak Ridge Leadership Computing Facility, which is a DOE Office of Science User Facility supported under Contract DE-AC05-00OR22725. This research also benefited from computational resources provided by National Energy Research Scientific Computing Center (NERSC). PNNL is operated by Battelle Memorial Institute for the United States Department of Energy under DOE Contract No. DE-AC05-76RL1830. NERSC is a U.S. Department of Energy Office of Science User Facility operated under Contract No. DE-AC02-05CH11231.

\bibliography{main}

\begin{widetext}
\vfill
\footnotesize

\framebox{\parbox{\textwidth}{
The submitted manuscript has been created by UChicago Argonne, LLC, Operator of Argonne National Laboratory (``Argonne''). Argonne, a U.S.\ Department of Energy Office of Science laboratory, operated under Contract No.\ DE-AC02-06CH11357 and Pacific Northwest National Laboratory (PNNL), a U.S.\ Department of Energy Office of Science laboratory, operated by Battelle Memorial Institute for the United States Department of Energy under Contract No DE-AC05-76RL1830. The U.S.\ Government retains for itself, and others acting on its behalf, a paid-up nonexclusive, irrevocable worldwide license in said article to reproduce, prepare derivative works, distribute copies to the public, and perform publicly and display publicly, by or on behalf of the Government.  The Department of Energy will provide public access to these results of federally sponsored research in accordance with the DOE Public Access Plan. 
http://energy.gov/downloads/doe-public-access-plan}}.
\end{widetext}
\end{document}